\documentclass[12pt]{article}
\pdfoutput=1
\usepackage{amssymb}
\usepackage{graphicx}
\usepackage{graphics}
\usepackage{amsmath}
\def\const{\mbox{const}}

\def\l{\left(}									
\def\r{\right)}

\newcommand{\be}{\begin{equation}}
\newcommand{\ee}{\end{equation}}
\newcommand{\ba}{\begin{align}}
\newcommand{\ea}{\end{align}}
\newcommand{\bg}{\begin{gather}}
\newcommand{\eg}{\end{gather}}
\newcommand{\bseq}{\begin{subequations}}
\newcommand{\eseq}{\end{subequations}}

\begin{document}
\title{Scalaron production in contracting astrophysical objects
}
\author{Dmitry Gorbunov$^{1,2}$, Anna Tokareva$^{1,3}$\\
\mbox{}$^{1}${\small\em Institute for Nuclear Research of Russian Academy of
  Sciences, 117312 Moscow,
  Russia}\\  
\mbox{}$^{2}${\small\em Moscow Institute of Physics and Technology, 
141700 Dolgoprudny, Russia}\\ 
\mbox{}$^{3}${\small\em Faculty of Physics of Moscow State
  University, 119991 Moscow, Russia}
}
\date{}

\maketitle

\begin{abstract} 
We study the creation of high energy SM particles in the Starobinsky
model of dark energy (a variant of $F(R)$-gravity) 
inside the regions contracting due to the Jeans
instability. In this modification of gravity the additional degree of freedom
-- scalaron -- behaves as a particle with mass depending on 
matter density. So when the mass changes light scalarons could be created at a
non-adiabatic stage. Later scalaron mass grows and could reach large
values, even the value of $10^{13}$ GeV, favored by early-time
inflation. 
Heavy
scalarons decay contributing to the cosmic ray flux. 
We analytically calculated the
number density of created particles for the case of exponential
(Jeans) contraction and found it negligibly small provided the
phenomenologically viable and cosmologically interesting range of
model parameters. We expect similar results for a generic model of
$F(R)$-gravity mimicking cosmological constant. 
\end{abstract}

\section{Introduction}

Numerous observational data requires the new
component in the right hand side of the Einstein equations called dark
energy which cause the accelerated expansion of the Universe. The
simplest and still viable candidate for the dark energy is obviously a
cosmological constant. But its unnaturally small value engenders
investigation of other ways to explain the observational
data.

$F(R)$-gravity provides a framework for
constructing models of dark energy with time dependent equation of
state $p/\rho\equiv w=w(t)$, moreover, at some stage one has $w<-1$ (
see \cite{Sotiriou:2008rp} for a
review). Such models of modified gravity may also explain an
inflationary stage of the early Universe providing the unified mechanism
to describe both stages of accelerated expansion.

The choice of function $F(R)$ is still phenomenological to a large
extent.  It must be self-consistent from the theoretical point of
view, explain the cosmological data and pass all solar system and
astrophysical tests. The natural question is how to distinguish
$F(R)$-gravity from other models of dark energy? 
The most straightforward way is to
improve the sensitivity of overall cosmological analysis to the dark
energy equation of state. However, apart of serious systematic
uncertainties there are physically motivated degeneracies in the
cosmological observables with respect to physical parameters. In
particular, specific effects of $F(R)$-gravity at small spatial scales
can be cancelled by massive (sterile) neutrino, which dynamics works
against modified gravity \cite{Motohashi:2012wc}.

An attractive idea of probing $F(R)$-gravity
\cite{Arbuzova:2012su} is associated with 
possible production of high energy particles (scalarons) 
in space regions where matter density
changes. It was claimed in Ref.\,\cite{Arbuzova:2012su} that the
growing curvature oscillations which decay to high energy particles
may lead to significant impact on the flux of ultra high energy cosmic
rays. This result is rather unexpected, because 
high frequency oscillations (i.e. heavy particles) 
are produced by a slow process (the
structure formation) very inefficiently. Moreover, cosmological
evolution naturally gives zero initial amplitude for such
oscillations, which can be associated only with the scalar mode 
(heavy scalaron).  

In this paper we study the processes of {\it quantum particle production} in
$F(R)$-gravity using the method of Bogolubov transformations. We
consider $F(R)$-gravity being equivalent to normal gravity with
additional scalar field (scalaron) having a complicated potential in
the Einstein frame. The form of potential depends on surrounding
density providing the scalaron to be a chameleon field. Scalaron is
very light at densities close to the present energy density of
the Universe and heavy at larger densities. It is known that the
particle with time-dependent mass may be created in a quantum theory
if the typical process rate is close to the mass value. 
Scalaron may be born being
light and then its mass may grow while the object contracts. When
scalaron becomes heavy it decays  to high energy Standard Model
particles.  
We consider the same $F(R)$-model as
in \cite{Arbuzova:2012su}, where such processes can be investigated
analytically and obtain that in opposite to \cite{Arbuzova:2012su}
the number density of created particles is unfortunately too small to
be observed in all realistic contracting regions (astorphysical
objects) in the Universe.

The paper is organized as follows. In Section 2 we describe the
construction of function $F(R)$ which is appropriate both for the
inflation and dark energy. In Section 3 we introduce the Einstein frame
approach to the $F(R)$-gravity in which the additional scalar field
has a mass depending on the background energy density. In Section 4 we
calculate the number density of produced particles in contracting due
to the Jeans instability object and discuss the particle production
rate in different contracting regions of the Universe (astorphysical
objects). We conclude then in Section 5.

\section{Description of the model}

The present day acceleration of the Universe expansion can be
described in terms of $F(R)$-gravity by action \footnote{The
  metric signature is $(-+++)$.} \cite{Starobinsky:2007hu} 
\be
S=\frac{M_{\rm P}^2}{2}\int d^4x \sqrt{-g}F(R), 
\ee 
with $F(0)=0$
reflecting the disappearance of the cosmological constant in the
Minkowski flat space limit.

Any viable $F(R)$ function must obey the classical and quantum
stability conditions: $F'(R)>0$, $F''(R)>0$.  It was introduced to
mimic $\Lambda$CDM model in the late-time Universe, so in the limit of
small curvatures $F(R)\approx R-2\Lambda$ with dark energy density
$\rho_{\Lambda}=\Lambda M_P^2$.  Moreover, the
  second derivative of $F(R)$ must also be bounded from above, 
  $F''(R)<\const$ (see \cite{Appleby:2009uf} for details) 
to avoid early-time singularities at $R\rightarrow\infty$.  
The last requirement is easily
satisfied for any $F(R)$ 
after adding $R^2$-term. This term with specially selected
coefficient may also provide the Starobinsky inflation in the early
Universe \cite{starobinsky}.

An example of a function appropriate for the dark energy proposed by
Starobinsky is \cite{Starobinsky:2007hu} 
\begin{equation}
F(R)=R+\lambda R_0\left(\left(1+\frac{R^2}{R_0^2}\right)^{-n}-1\right).
\label{starobinsky0}
\end{equation}
In the regime $R\gg R_0$ one has $F(R)\approx R-\lambda R_0$, one has
providing a cosmological constant. Parameter $R_0$ fixes a scale which
corresponds to the dark energy density $\rho_{\Lambda}$ (it is valid
for $\lambda \gtrsim1$): 
\be
\label{Sec2-eq-add1}
R_0\equiv\frac{2}{\lambda}\frac{\rho_{\Lambda}}{M_P^2}\,.  
\ee
To avoid early-time singularity hereafter we  use 
function $F(R)$ with the $R^2$-term added:
\begin{equation}
F(R)=R+\lambda R_0\left(\left(1+\frac{R^2}{R_0^2}\right)^{-n}-1\right)+\frac{R^2}{6M^2}~.
\label{starobinsky1}
\end{equation}

As discussed above the last term in \eqref{starobinsky1}  
is also appropriate for the usual
Starobinsky inflation \cite{starobinsky} in the early Universe if we choose
$M=3\times10^{13}$\,GeV; an impact of the second term of
(\ref{starobinsky1}) is
negligible for corresponding large values of
curvatures. 

There is a problem (described in
\cite{Appleby:2009uf}) with the subsequent stage of scalaron
oscillations. At this stage zero and even negative values of $R$ may
be obtained. Then it is easy to see that later the Universe 
unavoidably arrives 
at $F''(R)<0$ leading to the quantum instability of the theory. 
However
it is possible to construct a function $F(R)$ which provides similar
late-time cosmology but does not suffer from instabilities at
post-inflationary epoch. 
For example (here we neglect the presence of $R^2$-term) 
\be 
F''(R)\propto \frac{1}{1+(R/R_0)^{2n+2}} 
\ee 
lead to the similar to what one has with \eqref{starobinsky1} 
results at large curvatures. In what follows we 
use function (\ref{starobinsky1}) for simplicity and being
interested only in the late time Universe evolution. But we need to
ensure that the evolution of $R$ does not put its value to the region
of forbidden curvatures.

Starobinsky model \eqref{starobinsky1} 
has two free parameters: $\lambda$ and $n$ ($R_0$ is fixed 
by \eqref{Sec2-eq-add1}, while $M$ is fixed to explain the early-time
inflation). The stringent 
restriction on the value of $n$ follows from local gravity constraints
on the chameleon gravity \cite{Tsujikawa:2007xu}. It gives $n\gtrsim
1$. And parameter $\lambda$ is bounded only from the stability
condition for the de-Sitter minimum (see for a review
\cite{DeFelice:2010aj}). This bound mildly varies with $n$ as 
\begin{equation}
\label{inequality}
\lambda>f(n)\;\;\;\text{and}\;\;\;f(n)\approx n/(2n-4/3).
\end{equation}

\section{Einstein frame picture: scalaron as chameleon}

$F(R)$-gravity can be considered in the Einstein frame where it
describes usual Einstein gravity with an additional scalar field
(scalaron) coupled to the matter fields as chameleon field
\cite{Gannouji:2012iy}.  The scalaron potential is
\be
\label{Sec3-eq-add2}
V(\phi)=\frac{M_P^2}{2F'(R)^2}\l RF'(R)-F(R)\r ,
\ee
where $R=R(\phi)$ solves equation 
\be
\label{Sec3-eq-add3}
F'(R)=e^{\frac{\sqrt{2}\phi}{\sqrt{3}{M_P}}} .
\ee

Through the gravity interaction scalaron couples to all the matter
fields effectively described in cosmological context 
as ideal fluid with energy
density $\rho$ and pressure $p$. This coupling modifies its potential 
\cite{Gannouji:2012iy}:
\be
\label{Sec3-eq-add1}
V_{eff}(\phi)= V(\phi)+\frac{\rho- 3 p}{4} e^{-4\bar{\phi}} 
\ee
The minimum $\phi_{min}$ of $V_{eff}$ can be obtained by putting into 
\eqref{Sec3-eq-add3} the solution $R_{min}$ of equation
\be
2F(R)-RF'(R)=\frac{\rho- 3 p}{M_P^2}.
\ee
with $\phi_{min}$ written in terms of $R_{min}$.

For (\ref{starobinsky1}) we can approximate the solution as 
(the greater $\lambda$, the better the accuracy)
\be
R_{min}\simeq (4+\tau)\frac{\rho_\Lambda}{M_P^2}, 
~~~~\bar{\phi}_{min}=\frac{\sqrt{3}M_P}{\sqrt{2}}\log(F'(R_{min}))
\ee
where 
\[
\tau\equiv(\rho-3 p)/\rho_{\Lambda}\,.
\]

The effective scalaron mass in this minimum is
\be 
\label{eq:mass1}
m_{eff}^2=\frac{1}{3F''(R_{min})}\l1-
\frac{R_{min}F''(R_{min})}{F'(R_{min})}\r\simeq \frac{1}{3F''(R_{min})}
\ee
Thus for model (\ref{starobinsky1}) we obtain the
scalaron mass which depends on the surrounding energy density and pressure:
\be 
\label{m_eff}
m_{eff}^2=\frac{M^2 m^2 }{M^2+m^2} ,
\ee
where
\be 
\label{mass_approx}
m^2=\frac{1}{12n(2n+1)}\l \frac{\lambda}{2}\r^{2n}\frac{\rho_{\Lambda}}{M_P^2}(4+\tau)^{2n+2}~.
\ee
In a particular range of densities $m_{eff}$ strongly depends on
$\tau$. Obviously (see eqs.\,\eqref{mass_approx}, \eqref{m_eff}) $m$
is small there, $m<M$, leading to
\be 
\label{star_cond}
4<\tau < \l \frac{M
  M_P}{\sqrt{\rho_{\Lambda}}}\r^{\frac{1}{n+1}}\l\frac{2}{\lambda}\r^{\frac{n}{n+1}}
\l 12n(2n+1)\r^{\frac{1}{n+1}}= 
\l 1.4\times
10^{55}\r^{\frac{1}{n+1}}\l\frac{2}{\lambda}\r^{\frac{n}{n+1}} 
\l 12n(2n+1)\r^{\frac{1}{n+1}}~.
\ee
The situation when scalaron behaves as chameleon (i.e. inequality
\eqref{star_cond} is fulfilled) can be realized in 
a large variety of astrophysical objects at different scales depending
mostly on a choice of $n$. When mass of the particle varies with 
changing surrounding density such particle can be created if the
adiabaticity condition is violated. In the next section we calculate
the corresponding number of created particles.

The scalaron is unstable because it is coupled to all non-conformal
fields that are presented in the matter lagrangian. If we consider the
Standard model of particle physics 
to describe all the matter content then the scalaron
presumably decays to Higgs bosons (if kinematically allowed) 
\cite{Gorbunov:2010bn} with decay
rate 
\be 
\Gamma=\frac{m_{eff}^3}{48\pi \,M_P^2}\,, 
\ee
Higgs bosons decay immediately producing 
a flux of protons, electrons, neutrino which provide a possibility to
find such events in a case of significant scalaron production.

\section{Particle production in contracting objects}
\subsection{Mathematical problem}

A particle with mass depending on the surrounding density can be
produced in contracting clouds when the adiabaticity condition is
violated. Consider an object contracting due to the Jeans instability
with $\rho(t)=\rho_0 e^{t/t_J}$ ($t>0$, with $t_J=M_P/\sqrt{\rho_0}$). In
order to calculate the number of produced particles we use the
standard approach of Bogolubov transformations described for example
in a textbook \cite{Birrell:1982ix}. According to this method we write
an equation for the scalaron mode function $\varphi$ with momentum
$k$:
\be
\label{mode}
\ddot{\varphi}+(k^2+m_0^2\,e^{2\beta t})\varphi=0 \ee where the
definitions are \be m_0\equiv
\frac{\alpha^{n+1}}{\sqrt{12n(2n+1)}}\l\frac{\lambda}{2}\r^n\frac{\sqrt{\rho_{\Lambda}}}{M_P},~~~~
\alpha\equiv\frac{\rho_0}{\rho_{\Lambda}}, ~~~~\beta\equiv
\frac{(n+1)}{t_J}.  
\ee 
For the chosen $\rho(t)$ the contraction
starts at $t=0$, so we postulate vacuum initial conditions
$\varphi=1/\sqrt{2\omega}$, $\dot{\varphi}=-i\omega\varphi$ (where
$\omega=\sqrt{k^2+m_0^2\,e^{2\beta t}}$), at $t=-\infty$. Since
$\rho=\const$ for $t<0$ so such conditions may be set for any moment
$t<0$, e.g., set them at $t\rightarrow (-0)$ -- just before the
contraction starts. The adiabaticity condition
\cite{Birrell:1982ix} \footnote{Usually it is equivalent to the 
condition $\dot{\omega}/\omega^2\ll 1$ which means that mass of the
  particle must exceed the characteristic rate of the
  corresponding process.}
\be 
\label{ad-cond}
\left|\frac{\ddot{\omega}}{\omega^3}-\frac{3}{2}\l\frac{\dot{\omega}}{\omega^2}\r^2 \right|\ll 1
\ee
is violated only for $t<1/\beta$ so after this
time particles are not produced.

Equation \eqref{mode} with vacuum initial conditions 
can be analytically solved in terms of Bessel
functions. An exact form of the relevant Bogolubov coefficient is 
(up to an irrelevant complex phase) 
\be
B=e^{\frac{\pi  k}{2\beta}}\,
\frac{\sqrt{\pi}}{ 2 \sqrt{2\beta}}\,
\sqrt[4]{m_0^2+k^2}
\,  
\l
\frac{i m_0 \l H_{i \frac{k}{\beta}+1}^{(2)}(\frac{m_0}{\beta})
-H_{i \frac{k}{\beta}-1}^{(2)}(\frac{m_0}{\beta})\r
}
{2 \sqrt{m_0^2+k^2}}
+
H_{i  \frac{k}{\beta}}^{(2)}(\frac{m_0}{\beta})
\r
\,,
\ee
where $H^{(2)}_a(x)$ is the Hankel function.

Let us first discuss the case of $m_0<\beta$. It corresponds to the
situation where the scalaron is created being light at production
($m \sim \beta$) and a
bit later, when adiabaticity condition \eqref{ad-cond} gets restored. 
After that its mass grows until the moment when
scalaron decays to SM particles as discussed in the previous section.
The number of created scalarons can
be numerically obtained in this case as:
\be
\label{beta}
n_p=\frac{4\pi}{(2\pi)^3}\int k^2\,|B|^2\,dk\simeq C \beta^3~,
\ee
where $C=4.9\times 10^{-4}$.

In the opposite case $m_0>\beta$ one expects that particle
production is suppressed because massive particles cannot be created
in a slow process. But numerically we obtain
\be
\label{m_0}
n_p=\frac{4\pi}{(2\pi)^3}\int k^2\,|B|^2\,dk=C m_0 \beta^2~
\ee
with $C=6.23\times10^{-3}$. It looks strange that the larger the mass
the more particles are produced. The reason is connected with the fact
that the mass depends on time in a non-smooth way in the simple
mathematical model that we considered. Particles are produced mostly
at the time close to $t=0$ where the mass dependence on time is not
smooth. Divergent second order time derivative of  
$\omega$ leads to violation of
adiabaticity condition \eqref{ad-cond} at $t=0$. But actually it is natural to expect that the
contraction starts in a smooth way with the typical time $t_0>L>t_J$
(for causality reasons) where $L$ is a size of object. If we use such
kind of smoothing we obtain an expected suppression of particle
production because in this case modes evolve adiabatically.

In order to illustrate such a suppression we can consider a smooth dependence $m^2(t)=m^2_0(1+e^{2\beta t})$ with $m_0>\beta$ and put vacuum initial conditions at $t=-\infty$. After that one can analytically solve the equation like \eqref{mode} and obtain the Bogolubov coefficient to be (up to an irrelevant phase)
\be 
B=\frac{1}{\sqrt{2\pi q}} |\Gamma(1-i q)| e^{-\frac{\pi q}{2}},~~q=\frac{\sqrt{m_0^2+k^2}}{\beta}
\ee 
In the limit of large $q$ one finds 
$B\simeq e^{-\pi q}$. For $m_0>\beta$ we may write the produced number density as 
\begin{eqnarray}
n_p=\frac{4\pi}{(2\pi)^3}\int k^2\,|B|^2\,dk=\frac{4\pi \beta^3}{(2\pi)^3}\int^{\infty}_{m_0/\beta} |B(q)|^2\sqrt{q^2-(m_0/\beta)^2}\,q\, dq\approx \\ \approx \frac{4\pi \beta^3}{(2\pi)^3}\int^{\infty}_{m_0/\beta} e^{-2\pi q}\sqrt{q^2-(m_0/\beta)^2}\,q\, dq\sim \beta^3 e^{-\frac{2 \pi m_0}{\beta}}.
\label{theorem}
\end{eqnarray}
We see that in a realistic
model particle production is exponentially 
suppressed in a case of $m_0>\beta$. 
That is in accordance with the Rubakov theorem: 
{\it If you do everything correctly, the result is correct}. 
We can 
not calculate the exact number of produced particles in a
model-independent way, because it 
depends on details of the Jeans instability development. 

The result \eqref{beta} for $m_0<\beta$ is still correct because
particles are mostly produced not at $t\simeq 0$ but at the moment
when $m\sim\beta$, mass dependence on time is smooth and our
approximation works. Also it can be proved that the number of produced
particles does not depend on the details of how the Jeans instability
starts to evolve.

\subsection{Applications}

\subsubsection{Cosmic structure formation}

Provided the inequality \eqref{inequality} 
the condition $m_0<\beta$ is satisfied only for initial densities
$\alpha=\rho/\rho_{\Lambda}<8$ for the viable region of model 
parameters which corresponds to recent and
ongoing structure formation processes. Using (\ref{beta}) we can
estimate the 
number of created particles inside a region of typical size $L\sim 1$ Mpc:
\be 
N=n_p L^3=5\times 10^{-4} \beta^3 (1~{\rm Mpc})^3\lesssim 10^{-12}\,(n+1)^3.
\ee

We can see that only a negligible number of scalarons can be created
in the structure formation process. Only the enormous value of
  $n$ may lead to the noticable production which we disregard. 
As discussed before when $m_0>\beta$ particle production is strongly
suppressed so the earlier structure formation 
(star and galaxy formations) gives much smaller impact.

\subsubsection{Star formation in our galaxy}

The density range of contracting clouds that now form stars in our
Galaxy corresponds to the case $m_0\gg\beta$. As discussed before we
can not correctly describe the particle creation process because we
need to know how the Jeans instability evolves in details. But we can
obtain an upper limit on the number density of created particles 
as $n_p\sim\beta^3$ (in reality $n_p$ is much smaller because of
exponential suppression \eqref{theorem}) 
and calculate the corresponding flux of high
energy particles:   
\be 
F=\frac{n_p L^3}{r^2 t_J} N\sim 3\,(n+1)^3 \times
10^{-78}~{\rm cm^{-2}s^{-1}} 
\label{flux}
\ee 
Here $L=c_s\,t_J$ is a cloud size, $c_s\sim \sqrt{T/m_p}$ is a sound
speed in the gas (here we take a temperature $t=10$ K and $m_p$ to be a
hydrogen molecular mass), $r=10$ kpc is a characteristic distance in
  our galaxy and $N$ is a full number of objects that may be obtained
  from the known star formation rate of $3~M_{\odot}$ per year
  \cite{Allen}.

The measured flux of ultra high energy cosmic rays even at energy
$10^{20}$ eV, $F\sim
10^{-21}~{\rm cm^{-2}s^{-1}}$ \cite{Beringer:1900zz}, 
is many orders greater than the obtained number \eqref{flux}. 
So in any case the scalaron creation has 
a negligible effect in astrophysics.

\subsubsection{Expanding Universe}

Expansion of the Universe obviously implies changing energy
density. Scalarons are expected to be created at the moment when
$m_{eff}\sim H$ ($H$ is a Hubble parameter), hence from dimensional
analysis the number of created
particles can be roughly estimated as $n_p\sim H^3$, see also 
\cite{Zeldovich}. One can see that there were two moments in the
past of the Universe when $m_{eff}$ is close to $H$. The first moment
correspond to the period just after inflation. Scalarons created at
that time have decayed in the very early Universe and do not affect
the present Universe. The second moment, if exists, (for large
$\lambda$ scalaron mass $m_{eff}$ is always greater than $H$) is very
close to the present moment corresponding to the redshift $z<0.2$. So
we expect that the number density of created scalarons to be $n_p\sim
H_0^3$ which means that there is only one particle inside the present
horizon or even less.

\section{Conclusions and discussion}

We studied particle production in media with changing density which
takes place in $F(R)$-gravity or other chameleon models of the dark
energy. We performed a calculation of scalaron creation in the
Einstein frame for the Starobinsky dark energy model
\cite{Starobinsky:2007hu}. In the case of the exponential contraction
due to the Jeans instability the corresponding equation has an
analytical solution. Imposing vacuum initial conditions at the moment
when contraction starts we calculated the Bogolubov coefficient and
the number density of created particles. In all the realistic
situations the scalaron production is very inefficient.  Always less
than one particle inside the corresponding Jeans volume is
produced. Subsequent scalaron decay contirbution to the cosmic ray
flux is found to be infinitesimal.  In the Starobinsky variant of
$F(R)$-gravity considered here, formally, the production rate
increases with parameter $n$. However the latter must be enormously
huge to make the noticable effect. That implies drastic change in
$F(R)$-function at the particular value of curvature ($R=R_0$) which
is very unnatural choice with no grounds.

We thank F.\,Bezrukov, A.\,Dolgov, K.\,Postnov and S.\,Sibiryakov for
discussions. The work has
been supported by Russian Science Foundation grant 14-12-01430.


\end{document}